\documentclass[12pt,twoside]{article}
\usepackage{psfig}
\newcommand{\VolumeHeader}{}
\newcommand{\VolumeSerial}{}

\newcommand{\beq}{\begin{equation}}
\newcommand{\eeq}{\end{equation}}
\newcommand{\bea}{\begin{eqnarray}}
\newcommand{\eea}{\end{eqnarray}}

\newcommand{\LectureHeader}{Searching for...}

\begin{document}
\pagestyle{myheadings}
\markboth{\LectureHeader}{\VolumeHeader}
\markright{\VolumeHeader}


\begin{titlepage}


\title{Searching for continuous gravitational wave signals \\
- {\it The hierarchical Hough transform algorithm} -} 

\author{M.A. Papa$^\dagger$, B.F. Schutz$^\dagger$ and 
A.M. Sintes$^{\dagger}$
\\[1cm]
{\normalsize
{\it $^\dagger$ Albert Einstein Institut, Am Muehlenberg, 1 - 14476 Golm bei Potsdam, Germany.}}
\\[10cm]
{\normalsize {\it Lecture given at the workshop: }}
\\
\ Gravitational Waves: A Challenge to Theoretical Astrophysics
\\
\ 5-9 June 2000 
\\[1cm]
{\small \VolumeSerial} 
}
\date{}
\maketitle
\thispagestyle{empty}
\end{titlepage}

\baselineskip=14pt
\newpage
\thispagestyle{empty}


\begin{abstract}

It is well known that matched filtering techniques cannot be applied for searching extensive 
parameter space volumes for continuous gravitational wave signals. This is the reason why 
alternative strategies are being pursued. Hierarchical strategies are best at investigating 
a large parameter space when there exist computational power constraints. Algorithms of this kind are 
being implemented by all the groups that are developing software for analyzing the 
data of the gravitational wave detectors that will come online in the next years. In this 
talk I will report about the hierarchical Hough transform method that the GEO 600 data 
analysis team at the Albert Einstein Institute is developing. 

The three step hierarchical algorithm has been described elsewhere \cite{Papa99}. In this talk I will 
focus on some of the implementational aspects we are currently concerned with.

\end{abstract}

\vspace{6cm}




\newpage
\thispagestyle{empty}
\tableofcontents

\newpage
\setcounter{page}{1}


 

\section{Introduction}

It is expected that spinning neutron stars, whether isolated or in a binary system, emit continuous
gravitational waves. For example, they emit if their shape presents some deviations from axisymmetry. The signals are though expected to be weak: for example from the measured spin-down rate of the period of the Crab or of the Vela pulsars one can place an upper limit to the waves amplitude of $\sim 10^{24}$. For the Crab pulsar this is a factor of a few higher than the $99\%$ confidence limit for a coherent $10^7$ s search with initial LIGO. In general, 
a coherent integration for extended periods of time - of the order of 
a few months - is necessary in order to accumulate signal
to noise sufficient for detection. However, if the
phase evolution of the signal is not known in advance, many coherent
searches are necessary, each for every signal template resolvable in
parameter space. As the time baseline of the coherent integration
grows, the number of templates that one must search over grows
dramatically and the computational requirements soon become prohibitive. It 
is then necessary to explore alternative search strategies \cite{Brady1998}. Hierarchical
algorithms are the best way to explore large parameter space volumes
with limited computational resources. In this talk I shall present the 
hierarchical algorithm that is being developed and implemented
at the Albert Einstein Institute to search for continuous gravitational 
wave signals emitted by isolated sources. Somewhat similar strategies are also being pursued by the
Virgo Rome group for the VIRGO experiment and by the UWM (University of Wisconsin, Milwaukee) group for the LIGO
experiment. These approaches share many common features and the AEI
group is actively collaborating with both the Virgo-Rome group and the 
UWM group.

\section{The hierarchical Hough transform search strategy}

The search strategy consists of three main steps: 
\begin{itemize}
\item
a first coherent search over data segments lasting about a day - it is foreseen that the total 
observation time will be of the order of a few (one to four) months.
\item An incoherent search over the total observation time.
\item 
Finally, again a coherent search with a time-baseline that is a sizeable fraction of the 
total observation time. 
\end{itemize}

The latter search is limited to the regions in parameter space that
have been selected in the previous stage. This is why hierarchical strategies are so convenient at 
investigating large parameter volumes: in these circumstances the minimum signal detectable with a 
given confidence is larger than that one could detect with a priori information on the source 
parameters. Thus, one does not loose in sensitivity by using in the first stages of the search 
less sensitive and more computationally affordable methods. With these one can identify which 
are the regions in parameter space that it is worth investigating with a higher sensitivity \cite{Brady00}.

The procedure has been designed to be performed on a cluster of loosely coupled processors. 
Each node will run the same code and search a different region in parameter space. As a matter
of fact, each node will search a different intrinsic emission frequency band. The signal 
parameters are: intrinsic emission frequency, position in the sky and a set of 
spin-down parameters. The maximum order of the spin-down parameters depends on the class of sources 
that one is investigating. The minimum spin down age of a class of sources embodies 
all the information necessary to define how many spin down parameters are necessary to 
describe the frequency evolution of the system.

Distributing the computational load with respect to the intrinsic frequency search parameter 
is convenient because it allows a natural distribution of the data set among the nodes. This is 
immediately understood if one reasons in the frequency domain. The maximum Doppler shift that a
 monochromatic signal will be subject to, with respect to the emission frequency 
$f_0$, is $\sim 10^{-4}f_0$. This is the order of magnitude of the maximum frequency 
band of the data around $f_0$ that we need to access to search for a signal emitting at that 
frequency. The techniques that we have developed 
for our search make it possible to take advantage of this fact and distribute the computational 
load among the different processors according to the parameter intrinsic frequency {\it{and}} by 
distributing to the processors different data sets. This reduces some potential throughput problems since every node must 
access, at any time, only a small fraction of the total data set. For a 4 months search over a band 
of 500 Hz the total data that has to be accessible by any node is between 400MB and 4GB, depending 
on the total number of nodes (400MB refers to 100 nodes, 4GB to 10 nodes). HDs of few tens of GBs are completely standard.

\section{The coherent stage}

In this stage longer time-baseline Fourier transforms (DeFTs, hereafter) are produced from a set of shorter time-baseline Fourier transforms (SFTs, hereafter). At the same time  
a demodulation procedure takes place so that, if a signal having the appropriate parameters were 
present in the data, its demodulated spectra would present a sharp peak whose amplitude would be no more than $\sim 5\%$ lower than the total signal power. By appropriate signal parameters we mean 
here parameters that match those that define the phase evolution used in the 
demodulation procedure. In this case the frequency at which the peak appears is the intrinsic 
emission frequency of the signal: the procedure has removed the effects of the Doppler frequency 
modulation and of the intrinsic frequency evolution (spin-down).

\begin{figure}[h]
\centerline{\psfig{file=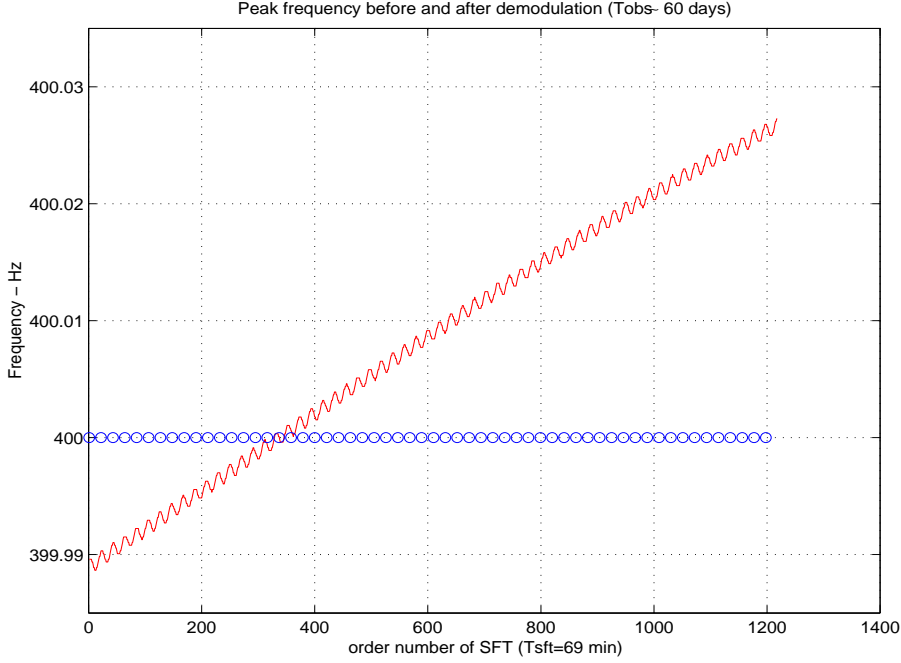,width=12.0cm,height=9.5cm,clip=}}
\caption{The continuous line shows the position of the instantaneous frequency of a signal coming from a position in the sky with ecliptic coordinates $(\beta,\lambda)=(0.0^o,0.0^o)$ and no spin-down parameters. The time-baseline of the SFTs is $\sim 1$ hour. The circles show the position of the instantaneous frequency after the demodulation procedure. The time-baseline of the DeFTs is $\sim 1$ day. The intrinsic emission frequency is $400$ Hz. Perfect match between template source and signal source was supposed.}  
\label{fig1}
\end{figure}

The continuous wavy line in fig. (\ref{fig1}) shows the position of a set of spectral peaks, as a function of time. These are the peaks of the 
signal in a set of SFTs. 
The circles in the same figure show 
the positions of the peaks in the demodulated spectra. Clearly the number of peaks has diminished by 
a factor $T_{DeFT}/T_{SFT}$, where $T_{DeFT}$ is the time-baseline of the DeFTs and 
$T_{SFT}$ the time-baseline of the SFTs. In this case $T_{DeFT}/T_{SFT}=21$.

There are a number of ways to perform this demodulation, which is a well known technique, for example 
in the field of radio astronomy. We have developed a method that combines only a few data points from 
each SFT in order to produce the demodulated one. The method has been described in \cite{Schutz1998} and \cite{Williams99}. The performance of this method is comparable to that of other methods (such as stroboscopic resampling, \cite{SCHUTZ1991a}) but it has the 
advantage of dealing with a much smaller data set. In particular, if the maximum expected shift of the 
instantaneous frequency during $T_{SFT}$ amounts to less than $1\over {2 T_{SFT}}$, one can only 
consider about 16 data points per SFT and loose no more than $5\%$ in signal power.

The DeFT in the $b$-th frequency bin, $\hat{x}_b$, is computed as follows:
\beq
\hat {x}_b=\sum_{\alpha=0}^{M-1}\sum_{k=0}^{N-1}\tilde x_{\alpha\beta} {1\over N} \sum_{j=0}^{N-1} 
e^{-2\pi i (\Phi_{\alpha j b}-{jk\over N})}.
\label{DeFT1}
\eeq
With $\tilde x_{ak}$ we indicate the $k$-th frequency bin of the FFT of the $\alpha$-th data set. There are $M$ such SFTs, each with $N$ frequency bins. $0<j<N-1$ is the time stamp of the short time-baseline data chunks. $\Phi_{\alpha jb}$ is the expected phase at time 
$t=N\alpha+j$ for an  
intrinsic emission frequency $b\over T_{DeFT}$. $\Phi$ parametrically depends on the 
source location and on the spin-down parameter values of the template source for which one is 
demodulating. If the phase evolution can be described as linear in $t$ during the time duration 
$T_{SFT}$, so that a linear Taylor expansion of $\Phi$ around the mid time point of every SFT time data chunk can be used, for large values of $N$
the summation over $j$ in eq. (\ref{DeFT1}) can be performed and eq. (\ref{DeFT1}) can be rewritten as
\beq
\hat x_b=\sum_{\alpha=0}^{M-1}e^{i y_\alpha}\sum_{k=0}^{N-1}\tilde x_{\alpha\beta} P_{\alpha k}(b),
\label{DeFT2} 
\eeq
with
\bea
P_{\alpha k}(b)={\sin{x'}\over x'}-i{1-\cos{x'}\over x'}\\
x'=\sum_{s} f_ss B_{s\alpha} - k\\
y_\alpha=\sum_{s} f_s A_{s\alpha}.
\eea
In the previous expressions $f_s$ indicate the spin-down parameters of the different orders 
(labeled 
by the index $s$\footnote{Note that when $s=0$ this quantity denotes the intrinsic frequency, thus 
has to be computed for the value corresponding to the index $b$.}), $A_{s\alpha}$ and $B_{s\alpha}$ 
are functions that depend on the phase evolution and so their value depends on $\alpha$ and on the 
spin-down parameters. The function $P_{\alpha k}$ is peaked around $x'=0$. Thus in the summation 
over $k$ in eq. (\ref{DeFT2}) one only needs to consider a few values (NTERMS) of $k$ around $k^*$ such 
that $x'(k^*)=0$. Eq. (\ref{DeFT2}) can then be  rewritten as
\beq
\hat x_b=\sum_{\alpha=0}^{M-1}e^{i y_\alpha}\sum_{k=k^*\pm NTERMS} \tilde x_{\alpha\beta} 
P_{\alpha k}(b).
\eeq
As said before, if $NTERMS$ is 8 the power loss due to this approximation is less than $\sim 5\%$. 
In principle then, to compute a DeFT for a single intrinsic frequency using $M$ SFTs, only
$64\cdot M$ bytes of data are needed. If  
$T_{DeFT}=1$ day and $T_{SFT}=1$ hour, then this amounts to less than $1.6$ kbytes. The point is that in the frequency domain the useful data 
for a given demodulation are localized and thus one can work with those only and forget about the 
rest of the data set. This is more involved to implement if one works in the time domain where the 
information is distributed across the entire data set.

\section{The incoherent stage}

This stage of the search works on spectra and not on Fourier transforms, and this is why it is called 
``incoherent''. With it, we piece together the information from {\it all} the demodulated spectra 
and select the regions in parameter space that it is worth inspecting coherently with a longer 
time-baseline than $T_{DeFT}$. The way this is done is by building an histogram in parameter space: 
high clustering in any pixel in such space indicates that there is suspect consistency of the data 
set with what would be expected if a signal were present in the data having the parameters 
associated with that pixel. 

The transformation that we use to go from the spectra to the histogram is the Hough transform. The 
basic idea of the Hough transform is very simple yet ingenious \cite{hough62}, \cite{Leavers93}. Suppose that one wanted to find a 
linear trajectory in an ensemble of points lying in the $x,y$ plane and wanted to estimate the 
parameters of the trajectory. Any such linear trajectory can be described by an equation of the kind 
$y=ax+b$. Uncertainties on $x$ and $y$ translate in gridding the parameter space $a$ and $b$ - 
suppose there is $N_a$ and $N_b$ values of $a$ and $b$ respectively in the resulting grid. For 
every value of $a$ and for every data point $x,y$ it is possible to derive a value of $b$: the 
number count in the corresponding $a,b$ pixel in the parameter plane should then be enhanced by
 unity. This is how one would perform the Hough transform in this simple case. To understand what the 
outcome would be, suppose that the only data points were on a line. In the histogram each data point 
would contribute 
a count in the right pixel and a count in other random $N_a$ pixels, different from data point to 
data point. So, after having transformed all the data points, we would find that the average count in any
 pixel is ${N_p\over N_b}$, whereas in the signal pixel it is $N_p$. Thus, we would observe a 
significant clustering in the pixel corresponding to the line parameters. If in the data set there 
were points that did not belong to the line, i.e. noise, these would raise the number count uniformly 
in the histogram, with the effect of diminishing the significance of the clustering level in the signal
 parameter pixel.

In our case the $x,y$ plane is a time-frequency plane. The points that appear in it correspond to the 
frequency bins in the spectra where a significant peak\footnote{We have developed techniques to 
define what a ``significant peak'' is. We do not want to go into this detail now, though. It suffices 
for the purposes of this presentation to know that a sensible way to define a significant peak 
exists.} has been identified. If a signal 
is present in the data, due to the residual modulation, the signal peak will appear displaced from the 
intrinsic frequency bin with a pattern in time that depends on the mismatch between the trial signal 
parameters used in the demodulation and the actual signal parameters. Identifying suspect peak 
patterns that are consistent with what we would expect from a putative source is the goal of this 
stage of the procedure. 

Suppose that the demodulation has been performed for a given sky position $\hat N$ and spin-down parameters $F_s$ and that a source is located at a different position $\hat n$ and that it has a different set of spin-down parameters $f_s$. The equation that governs the Hough transform in this general case can be cast in the following form:
\beq
\nu-F_0=\vec{\xi}(t)\cdot (\hat n - \hat N),
\label{HTeq1}
\eeq
with:
\begin{itemize}
\item $\nu$ the measured frequency at time $t$ - namely the frequency of the peak registered at time $t$
\item $f_0$ the intrinsic frequency of the signal at a time $t_0$ (usually the beginning of the observation)
\item $F_0=f_0+\sum_s \Delta f_s [T_{\hat N}(t)-T_{\hat N}(t_0)]^s$
\item $\Delta f_s=f_s-F_s$, the residual spin-down 
\item $T_{\hat N}(t)$, the time at the solar system barycenter for the sky position defined by the
 versor $\hat N$
\item $\vec{\xi}(t)=\{F_0+\sum_sF_s [T_{\hat N}(t)-T_{\hat N}(t_0)]^s\}{\vec {v}(t)\over c}+
\{ \sum_s s F_s[T_{\hat N}(t)-T_{\hat N}(t_0)]^{s-1}  \}{\vec x(t)-\vec x(t_0)\over c}$, a vector 
that depends on the demodulation parameters, on the position of the detector ($\vec x$) and on 
the relative source-detector velocity ($\vec v$).
\end{itemize}
For a given value of the intrinsic emission frequency and of the residual spin-down parameter $\Delta f_s$, 
eq. (\ref{HTeq1}) defines, for any given peak registered with a time stamp $t$, the locum of points
 were a source consistent with these parameters would be if it were to produce in the demodulated 
spectra labeled by the time stamp $t$ a peak at $\nu$. Such locum of points 
is derived by solving eq. (\ref{HTeq1}) with respect to $\hat n$. This results in a circle on the 
celestial sphere centered where the vector $\vec \xi (t)$ is pointing and with aperture angle 
$\cal\phi$ such that 
$\cos{\cal{\phi}}={\nu-F_0+\vec{\xi}(t)\cdot \hat N\over |\vec{\xi}(t)|}$.
For every registered peak it does not suffice to enhance the number count in the pixels in the 
histogram that satisfy eq. (\ref{HTeq1}). In fact, the registered frequencies all carry an uncertainty, 
due to the finite resolution of the spectrum. As a consequence the locum of points that is consistent 
with a given set of source parameters and a registered peak is not a circle but it lies 
inside an annulus, 
whose width depends on $F_0$, on the frequency resolution of the demodulated spectra and on the angle between $\vec\xi$ and $\hat N$. The 
borders of such annulus satisfy an equation like eq. (\ref{HTeq1}) with 
$\nu\rightarrow \nu+{1\over 2T_{DeFT}}$ and $\nu\rightarrow \nu-{1\over 2T_{DeFT}}$.

Direct solution of eq. (\ref{HTeq1}) involves, once all the variables are set, solving a linear  
equation in a trigonometric function of the ecliptic longitude, 
for every possible value of the other coordinate, the latitude. Thus, in this form, it 
is rather computationally costly. An alternative consists in reverting the argument and instead of 
transforming peaks into pixels, doing the contrary. In practice this means that given a pixel - 
defined by $\hat n$ - one computes $\cos\cal\phi$. This is easily done as a scalar product of the 
cartesian components of $\hat n$ and $\vec\xi$. Then one uses eq. (\ref{HTeq1}) to find what value of 
$\nu$ this corresponds to and then checks in the time-frequency diagram if a peak was registered at 
that frequency or not. If so, the number count can be enhanced in that pixel. The gain with respect 
to the direct solution for $\hat n$ 
is that one does not deal with inverse trigonometric functions. The draw-back is that one is systematically inspecting a large number of pixels that are 
not part of any annulus, at any  given time. A further gain can be obtained in these circumstances by 
a zooming approach. The concept is still that of the pixel-to-peak approach but one does not start by considering a pixel but rather a larger area. Thus it is possible to 
associate with this region not just a single frequency value but a range of frequencies. If peaks 
have been registered that lie in such frequency range then it means that there are annuli intersecting 
the region, otherwise it means that there are not. If there are no peaks associated with this 
area the procedure simply does not update the number count in that area and passes on to analyze 
another region. Note that the frequency interval associated with the area can be degenerate, namely 
reduce to a single frequency bin. In this case if a peak was registered at such frequency then the 
number count in all the pixels in the area can be enhanced by one and another region can be analyzed. 
If one is not in this degenerate case and if there are peaks associated with the region, then the 
procedure that we have just described is iterated on sub-patches inside the region. This process takes 
place up to a definite sub-patch size and from there on the pixel-by-pixel approach is used. 
Preliminary timing results of this method are encouraging when compared to the theoretical optimal 
performance of \cite{Papa99}, especially for the case where first order spin down parameters are included 
in the search. Alternative schemes are also under study. These comprise look-up table approaches, 
peak-to-pixel approaches with a simplified master equation valid for small regions around $\hat N$ 
and hybrid approaches where we try to take advantage of the positive features of the different 
methods while minimizing the drawbacks.

A different Hough histogram must be produced for every value of the intrinsic 
frequency and of the spin-down parameter residual. The ``brute force'' approach to this consists 
in scanning this parameter set and in producing for each of the parameter values a Hough histogram using all the peaks in the time-frequency diagram. 
But there exists an alternative way to do this.
Let us define a partial Hough map (PHM) as being a Hough
histogram generated by using peaks that come from the same
spectrum, given a value of the intrinsic frequency and supposing no spin-down residual. The 
corresponding {\it total Hough histogram is simply the sum of the PHMs} coming from the entire 
set of spectra. Moreover, if every partial
Hough map has been constructed by using the same intrinsic frequency, 
then the resulting total Hough map refers to the same spin down parameter value as the
one used in the demodulation (no residual, as said before) and that intrinsic frequency. But note 
that the effect of a non-zero residual spin-down in the signal is the same as having a time varying 
intrinsic frequency. This suggests a strategy to re-use PHMs computed for different $f_0$s at different 
times in order to compute a Hough histogram for a non-zero spin down residual. If one has thus a 
collection of PHMs referring to different intrinsic frequencies and different times one can construct 
the Hough map for any $f_0$ and spin-down residual value simply by adding together at different times 
PHM corresponding to different $f_0$s. In other words, by choosing in a suitable manner
the intrinsic frequencies of the different partial maps to be summed,
one can investigate the parameter space around the coherent
demodulation spin-down parameter value. This approach saves computations because it recognizes that 
the same sky locations can refer to different values of $f_0$ and spin-down residuals, and it avoids
 having to re-determine such sky locations more than once.

\section{Unresolved issues}

In principle there exist correlations in parameter space that allow considerable computational 
savings \cite{Brady1998}, \cite{Brady00}. Given a certain coherent demodulation time-baseline, for every time label it is 
possible to identify an optimal gridding of the parameter space. It is though not yet clear how 
much this gridding changes with time. Understanding this is crucial because the incoherent stage 
pieces together information from demodulated spectra with different time stamps. For this to work, 
the spectra must have been demodulated for the same physical parameters. Else, the physical parameter 
space region to be considered for the incoherent stage is the intersection of the different regions 
and an appropriate book-keeping must be set up in order to ultimately cover the whole parameter 
space while avoiding redundand demodulations or excessive storage requirements for the set of 
demodulated spectra. 

With time also the gridding of the parameter space for the incoherent stage changes. This means that 
the Hough transform pixel size in principle changes depending on what time stamp is attached to the 
peak that one is considering. In our simulations and operation count estimates we have always used 
the highest possible resolution, thus often over-resolving the parameter space and thus doing more 
computations than strictly necessary. The zooming approach adapts well to varying resolution grids, 
especially if the ``PHM strategy'' outlined in the previous paragraph is employed. This happens 
because each PHM refers to a single time-label, thus has its own resolution and the number count in
 each can be determined using that resolution, thus with the minimum number of operations. In order 
to ultimately sum together the PHMs, one can construct over-resolved maps from the lower resolution 
ones and obtain, in the end, the actual physical resolution over the total observing time in the 
different regions of the sky.

\section{Development plans}

The GEO detector is expected to have its first science data run in 'mid 2001. For that time the 
search code that performs the search for continuous signals must be in place. It is true that a 
latency exists between the time the first data are produced and the time the analysis is carried out. 
In fact, the total set must be acquired before the analysis is initiated. But we plan to use this time 
not for algorithm development but rather to understand the noise of the detector and refine our data preparation procedures. 

We aim at having a hierarchical code working on a single node by the beginning of 2001. As stated in 
the introduction, the procedure has been designed to run on a set of loosely coupled processors. We 
are not very concerned, thus, with the generalization of the single-node code to many machines. This 
is mainly a computational load distribution problem and, as far as we can see, there will be no 
reason to do this dynamically. Provided that the data preparation is effective, the intrinsic frequency band that each node has to analyze can be 
decided a priori and the data distributed accordingly once and for all. 

\section*{Acknowledgments}

The most complicated part of the algorithm is the efficient implementation of the Hough transform. We 
are developing this in collaboration with S. Frasca, F. Massaioli, C. Palomba and G. Amati at the 
University of Rome ``La Sapienza''. Part of the work presented here is fruit of our joint efforts. 
We also acknowledge useful discussions
 with our GEO colleagues A. Vecchio, C. Cutler and B. Sathyaprakash and our colleagues at UWM who are 
involved in the development of another hierarchical search algorithm, B. Allen, B. Brady, T. 
Creighton and B. Owen. We would like to thank S. Berukoff who is currently coding the coherent search algorithm, following the LAL (LIGO algorithm library) specifications and helping in the general LAL-compliant set-up of the whole procedure.


\end{document}